# Binaural recording methods with analysis on inter-aural time, level, and phase differences.

An Honours Thesis

submitted in partial fulfilment of the requirements for the degree of

Bachelor of Music with Honours in Audio Arts and Science

to

the Department of Audio Arts and Science

Yong Siew Toh Conservatory of Music

National University of Singapore

A/P. Zhou Xiaodong, Supervisor

by

Johann Tan Kay Ann

10 April 2020

This Honours Thesis represents my own work and due acknowledgement is given whenever information is derived from other sources. No part of this Honours Thesis has been or is being concurrently submitted for any other qualification at any other university.

Signed: ____JOHANN TAN____

# Acknowledgement

I would like to thank A/P Zhou Xiaodong for the guidance throughout my four years of undergraduate studies under his studio as well as for providing space and equipment to conduct the experiments involved in this study. I would also like to thank my peer, David for his support during the experimental conduct of this study as well as for help in documenting photographs used as illustrations in this thesis. Finally, I want to express my deepest gratitude to my family and closest friends for supporting me during the writing of this paper.

# Abstract

Binaural recordings are a form of stereophonic recording method that replicates how human ears perceive sound, these types of recordings create a 3D aural image around the listener and are extremely immersive when well recorded and listened to appropriately with headphones. It has wide applications in video, podcast, and gaming formats- allowing the listener to feel like they are there. Although binaural formats are seldom used for music applications, they have also been utilized in music ranging from Rock, Jazz, Acoustic, and Classical. In this paper, we will investigate the acoustical phenomenon that produces the binaural effect in audio recordings- including the ITD (Inter-aural time difference), the ILD (inter-aural level difference), IPD (inter-aural phase difference) as well as the monaural spectral difference that occurs between two ears so we can better understand the replication of human hearing in binaural recordings. Binaural recordings differ from regular stereophonic recordings as they are arranged in a specific way to account for HRTF (Head-related transfer function). The most common method of binaural recordings is with two high-quality omni-directional microphones affixed on a dummy head where the ears are located, although other methods exist without the use of a full dummy head.

# Contents

# 1. Introduction

The term 'binaural' was first coined by Alison (1858) in his paper on Stethoscope- relating binaural to the human ear and how our ears perceive sound in unique ways; just like how our two eyes can perceive depth, our ears have auditory depth perception (Grantham, 1995) which would not have been possible if we only had one ear. With a pair of ears, it allows us to localize sound, this 'spatial awareness' is a conscious knowing of where our physical self is within a space, whether in reality or a virtual environment. Spatial awareness is an awareness of how sound interacts with the acoustics of the physical environment; our bodies and how our heads create an acoustic shadow due to the head's baffle effect.

The use of binaural audio recording can turn playback listening into an immersive experience especially when it comes to soundscapes or ambiance recording which helps support the immersion of being in a space in which the binaural audio was recorded, this immersion is possible because of our human brain's extraordinary ability to localize sound which can be exploited using binaural playback- listeners can feel like they are physically there at the space of the recording or in a specific virtually crafted environment. The human ear's ability to localize sound has long been documented since the 18th century when the first experiment was carried out by Battista Venturi (1746–1822), positing that our ears can perceive minute differences in the sound input that we receive in both ears which our brains process in real-time. Venturi stated from his experiments that;

*"Therefore the inequality of the two impressions, which are perceived at the same time by both ears, determines the correct direction of the sound"* ([52], p. 186)

We now know what this '*inequality in two impressions*' entails, namely- the intensity, time, frequency, and phase differences of the sound inputs as they enter both our ears. In essence, the same 'inequality in two impressions' of both outputs in a binaural playback helps recreate the binaural effect when listeners play back a binaural recording. From this, the listeners can localise sound in the recording just as they would be able to in reality which contributes to an immersive listening experience. In this study, we will explore the differences in the inequality of sound that enters both ears and evaluate whether typical binaural recording methods can replicate these effects with a dummy head or Jecklin disc. Beyond this, we will also look at other binauralisation processes.

## 1.1 Background

In recent years, headphones have become more widely and readily accepted as a playback medium for audio content due to their affordability and popularity. Almost anyone today living in cities has a headphone lying around at home. With the rising popularity of headphone usage (a fundamental requirement of binaural audio playback) we would expect to see a rise in the relevance and listen-ability of binaural audio content and in it's regards to the growing popularity of immersive technology as a whole such as in virtual reality and augmented reality technology (VR/AR). (T Walton, 2017) A high-quality headphone is also not required for the enjoyment of binaural audio content, although it is recommended. The standard stereo loudspeakers will not be able to accurately play back binaural audio unless they have been tuned to optimize crosstalk cancellation between two speakers which is known as transauralization. In this paper, we will not consider the playback of binaural content through transaural stereo due to the incompatibility of binaural playback with loudspeaker usage. The challenges to transauralization are also comprehensive and can be another topic of its own.

Binaurally recorded content has many applications in media consumption. When used with headphones playback in video formats, it is more effective than stereo recordings for environmental sounds such as rain or cityscapes/soundscapes. (Rumori and Marentakis, 2017) One of the earlier films to incorporate binaural audio into their production (Bad Boy Bubby) won many film awards for its use of binaural microphones that were sewn onto the actor's wig. In VR/AR and First-person shooter (FPS) video game formats, binaural audio content has also been used creatively to great effect such as for horror elements in sound design like Hellblade and for dialogue (McGurk effect) as well as environmental sounds. In video games that require quick reactions like FPS games, binaural audio that is encoded into games can elevate the immersion. Approaching entities are usually heard first before they are seen (Bronkhorst 2000 Kidd et all 2005) [Army Research] as auditory stimuli reaction ($\mu = 284ms$) is faster than visual stimuli reaction. ($\mu = 331ms$) (Jose Shelton, 2010) We can expect that binaural audio will help cue gamers to react more quickly and accurately to what happens in-game. Binaural audio technology has recently become popular due to its use in ASMR content (Autonomous Sensory Meridian Response) produced by independent content creators and binaural audio is also becoming more prevalent in podcasting applications in order to add new dimensions to interviews and podcaster's perspectives. Other applications include ecological acoustic research for environmental noise/soundscape (ISO 12913-2:2018) as well as sound system optimization and assessment using an actual representation of how the public or concert-goers would be affected by concert hall acoustics using binaurally recorded content for evaluation instead of using single omnidirectional microphones which are not representative of how our ears would actually hear a sound. There are also applications in research relating to psychoacoustic.

It is important to note that binaural audio implementation in video or gaming formats is not simply the use of pan potentiometer or spatial stereo imaging processors in the mix whereby monaural recordings are panned to mimic sounds coming from the left, right, or back. While they do create the effect of sound sources coming from a specific direction, they are artificial and do not replicate the way our ears listen and thus are not true in the identification and directionality of sound sources in the way that binaural recordings are. The pan potentiometer only changes relative levels but does not include phase information, time differences, and sound reflections in physical spaces which takes the environment's acoustics into consideration. However, there have also been new developments in digital signal processing that allow the 'binauralization' of audio content by involving complex functions to create a binaural effect, such as the Anaglyph 3D VST which accounts for HRTF and allows for customization of head radius to replicate different 'listener' HRTF signature. Almost all collections of recordings ranging from music to films are considered to be 'artificial stereo' and only a handful of recordings are considered to be 'true stereo' which incorporates binaural recording and reproduction techniques.

## 1.2 Technology of Binaural Recordings

The word binaural consists of two words; bi- meaning two, and aural- meaning ears.

Binaural (two ears) and binaural audio initially simply referred to audio that involves playback of audio contents with two ears. Our modern understanding of 'binaural audio' has evolved over the years into what it is today, but it did not start as a technology meant for use in immersive media consumption. In fact, the first use of binaural technology dates back to 1881 when Clement Ader invented the 'Théâtrophone' which consisted of an array of 80 telephones that lined the edge of the stage in Opera Garnier theatre where paired microphone signals could be sent via telephone lines to subscribers so that they can listen to a performance in binaural audio while not being present in the theatre. A modified headset was used for this purpose and was the first ever dual-channel binaural stereophonic playback. Later, binaural technology was used in World War I by the Allies to detect and identify approaching enemy planes. These were large trumpet tubes that narrowed down and would fit onto a listener's ear. During World War 2, a similar device was made with hydrophones to detect submarines. Harvey Fletcher; well-known by the acoustics community who had contributed to the advances of hearing aid devices- used two microphones that were connected to a headphone in a binaural fashion. (Stephan Paul) However, all these previous inventions were not created for the 'recording' of audio content that represented how our pair of ears listens. It was only in 1930 that such devices started to be invented, first with artificial dummy heads and microphones positioned where the ears would be. Later, these 'mannequin' recording heads were improved by a handful of companies researching and developing such devices such as Bell Laboratories, Neumann, Sony, and Sennheiser, and in years to come, many more would follow. In the present day, binaural audio mainly refers to the recording and playback of audio content with two inputs or two outputs (L and R ear) to emulate the human ears. The use of binaural content has also moved towards immersive media consumption. Before we can discuss the different binaural microphone techniques that are popular today, it is useful to note the existing stereophonic microphone techniques that have contributed to the development of binaural microphone techniques today. These techniques include the AB technique, coincident stereo (XY, MS, and Blumlein), and near-coincident stereo (ORTF, NOS, and Faulkner). Whenever a second microphone is introduced to a recording setup, it will produce ITD and ILD as well as phase differences between the two microphones. Due to this, recording engineers sometimes have to reposition microphones to find the best stereophonic image in the context of what they are recording and also to avoid unwanted comb-filtering effects.

However, in the context of binaural recordings, the approach is different and it is desirable for sound engineers to find the best HRTF binaural image. Of all the existing stereophonic techniques, it is the near coincident technique; ORTF (Figure 1.1) that creates the most optimal quasi-binaural image. However, it is still considered to be a stereophonic recording method. In a quote by Francis Rumsey in 2001, he described ORTF;

*"One comprehensive subjective assessment of stereo microphone arrangements, performed at the University of Iowa (Cross, 1985), consistently resulted in the near- coincident pairs scoring amongst the top few performers for their sense of 'space' and realism." - Spatial Audio by Francis Rumsey, 2001*

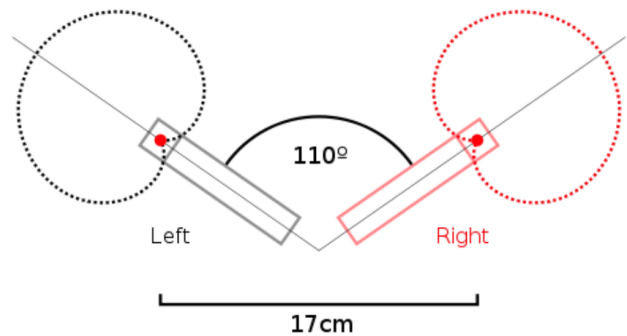

Figure 1.1: Example of ORTF Technique

### 1.2.1 Artificial Dummy Head Recording

Also known as 'Head and Torso Simulator' (Brüel & Kjær 4128D/C (Nærum, Denmark), the use of an artificial dummy head with microphones affixed on two sides of the 'dummy head' (see figure 8) is the most straightforward method for creating binaural recordings. It is not difficult to create a do-it-yourself (DIY) (Figure 1.2) binaural dummy head recording model or makeshift binaural dummy head using omnidirectional microphones and a manikin head. Still, it is possible to find ready-made binaural dummy head devices in professional audio where microphones are already affixed in place inside a dummy head with dual or single stereo output like the Neumann KU 100. (Figure 1.3)

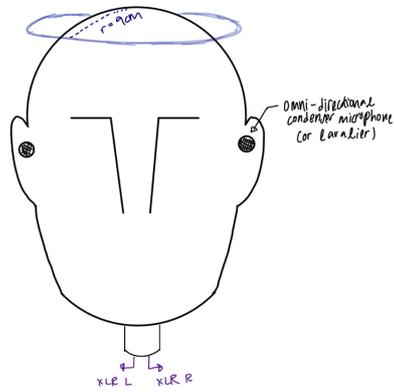

Figure 1.2: Typical drawings of DIY dummy binaural recording head example

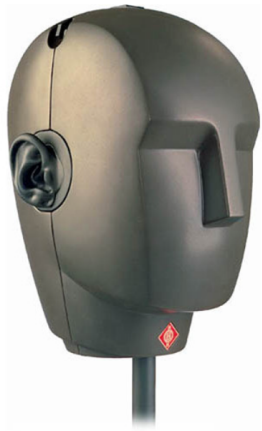

Figure 1.3: Neumann KU100, Picture taken from Neumann

Unless the full dummy binaural head includes the torso, the reflections from the torso will not be captured which affects frequency attenuations around 2-3kHz (Army Binaural)

### 1.2.2 Artificial Semi-sized Dummy Head Recording

However, because of the size and impracticality of carrying a full-sized dummy head microphone to field recordings, an alternative is to create semi-sized dummy head models to get similar results as a full-sized dummy head model for recording applications as it is much easier to carry around for field recordings. Albeit the binaural effect will not be as pronounced as a full-sized dummy head- compromising in the binaural quality for more practicality and convenience during field recordings. An example is the 3DIO binaural microphones (Figure 1.4) which include only the ears as an obstruction with a separation of about 190mm to represent ITD differences for signals input of both ears as well as an approximate of the head baffle effect which contributes to binaural effect.

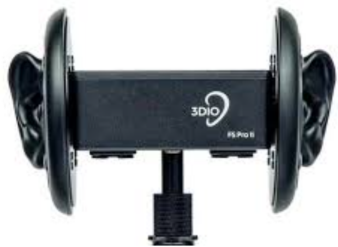

Figure 1.4: 3DIO Free Space Pro II, Picture taken from 3DIO

The binaural dummy head methods are very similar in their microphone positioning to the ORTF Stereo microphone technique mentioned above, which was a method devised around 1960 by the Office de Radiodiffusion Télévision Française (ORTF). The ORTF technique was used as a stereo recording technique using cardioid microphones at that time to roughly emulate the binaural sound of human hearing using ITD references and its level differences from the on and off-axis sound captured from the microphones being spread at 110 degrees and spaced 17cm apart. However, the technique is not as effective as the modern binaural recording technique today but is still extremely good for stereo recording due to the wide stereo image it captures and its compatibility for playback in loudspeaker formats which also makes ORTF more suitable in mono playback.

### 1.2.3   Baffled Stereo Pair Technique

Another group of recording methods used for binaural application is generally known as the baffled stereo technique where two of the same omnidirectional microphones are spaced facing away with an acoustic baffle in between- creating an acoustic 'shadow'. The acoustic baffle could be any material that has an acoustical absorption coefficient that is close to that of a human head. It would also need to be non-reflective such as melamine foam. This technique specifically exploits the attenuation of high frequencies from the acoustic baffle to reproduce the binaural effect and similarly to the previous dummy head recording method, the distance between two microphones forms an ITD reference that simulates how our ears listen as sound reaches both of our ears.

### 1.2.4   Sphere microphones

In some cases, spheres can be used as the acoustic baffle like that of the Schoeps Sphere (Figure 1.5) where a sphere material is used to emulate HRTF. However, for the Schoeps KFM 6 Sphere Microphone, its quality is in between that of a wide stereo image and a binaural image, while not completely being useful for a fully immersive binaural recording. Thus, like the ORTF stereo microphone technique, it is a better fit for loudspeaker playback and mono compatibility.

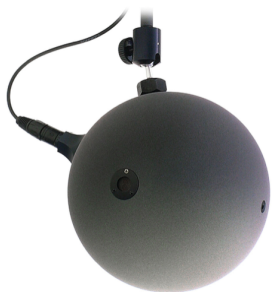

Figure 1.5: Schoeps KFM 6 Stereo Sphere Microphone, Picture taken from BHP

### 1.2.5 Jecklin Disc

The Jecklin disc or also called the Optimal Stereo Signal disc- developed by Jürg Jecklin (Figure 1.6) employs the use of an acoustic baffle in between two microphones spaced a distance that is farther apart than previous techniques (17.5cm) to create a quasi-binaural image. In this case, instead of being a small acoustic baffle material in between the omni microphones, a larger piece of material of around 35 cm- but less absorptive material such as plywood (or other woods), cork-board, foam, felt, or corrugated cardboard (12-inch thickness) is placed in between the microphones to create an acoustic 'shadow'. This technique is easy to make as the materials are readily available and only requires a good omnidirectional microphone. Commercial Jecklin Disc microphone products also exist made by Core Sound which only includes the disc and excludes the microphones so the sound engineer can experiment with different omnidirectional microphones for his setup. (Figure 1.7)

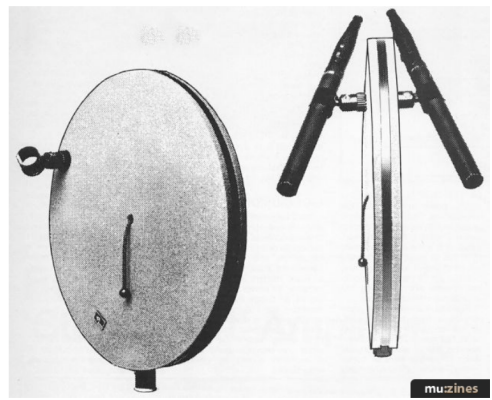

Figure 1.6: Jecklin Disc by Core Sound, Picture taken from mu:zines

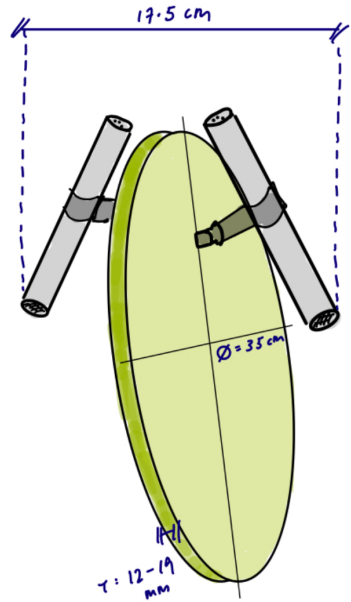

Figure 1.7: Typical drawings of DIY Binaural Jecklin Disc Setup w/ omni-direction microphone

### 1.2.6    Binaural headsets or in-ear microphones

In recent years, microphones have been made smaller that they can fit on headsets or in-ear devices that can be used as a peripheral to record as well as playback audio. Such recording devices are able to record good quality and accurate binaural images because they use the actual user's head as an acoustic baffle to create an individualized HRTF-based recording which would also involve the structure of our ear's pinnae. This is the most convenient and discreet method (as it is also completely hands-free) for field binaural recordings. There are professional binaural headset microphones available like the DPA 4560 CORE Binaural Microphone as well as consumer binaural headsets like the Ambeo Sennheiser, Hooke Audio, Lifelike Headset, and Roland CS-10. There is no single superior technique for binaural recordings and depends on the context and application used. For instance, a fully immersive experience for applications in VR would likely work the best with full or semi-dummy head binaural recordings whereas for the music and podcast application-a quasi-binaural effect may be more desirable so that it is compatible with loudspeaker playback. The sound engineer should consider what his applications are and try out a few iterations in positioning and adjustment to the standard binaural recording techniques to find the best method.

# 2. Theories & Concepts

## 2.1   Duplex Theory

As we have discussed earlier, binaural audio content allows us to localize sound based mainly on the subtle difference we hear in the time delay (ITD) and level differences (ILD) between two channels to our ears (Lord Rayleigh, 1907) except direct frontal or rear sounds which have no useful ITD or ILD references for sounds arriving from the front or rear. This is also named the Duplex Theory by Rayleigh. Further revisions to the Duplex Theory and contributions by Woodworth (1938) and Feddersen (1957) on human experimentation, found that the ITD as well as ILD is only effective on certain frequencies and our ears process ITD and ILD concurrently- working together in a complement across a wide frequency spectrum to help us localize sound sources. For the average radius of an adult human head which is typically within r = 8-9cm. The Duplex theory states that the inter-aural time difference (ITD) of a sound source on either an extreme far-left or extreme far-right (90 degrees from the front of the head) would roughly have a 0.6 ms delay between both ears and would be effective for sound localization in the low frequencies upwards to $f_i$ = 1500 Hz (Woodworth). The useful frequency for ITD localization will be slightly higher for larger head radiuses. Beyond the frequency, $f_i$ (>1500 Hz), it becomes difficult to localize sounds via ITD cues, although it would still be possible to tell the general direction in which the sound source is emitted. For more accurate localization, our ears rely on inter-aural level differences (ILD or IID) cues in the higher frequencies. The useful frequency for ILD localization starts from $f_L$ = 2000 Hz, However, localization by ILD is still not especially efficient for the frequencies ranging between 2000 to 4000 Hz and unreliable for the frequencies below 1500 Hz. This is because, at lower frequencies, the sound's wavelength is long and can bend around the head, causing insignificant differences in ILD which creates difficulty in exact sound localization. For individual HRTF, the ILD will not have any significance below the minimum frequency, $f_{Lmin}$ where the individual's head diameter is 1/3 of the frequency's wavelength. (Howard

& Angus, 1996)

The ILD is also caused by many other secondary and tertiary factors, these secondary factors include sound energy loss from traveling the distance between one ear to the other, as well as the absorption of sound energy by the human head which acts as an acoustic shadow. Furthermore, tertiary factors for sounds arriving from the rear of our heads involve our ear's pinnae which acts as an obstacle and spectral modifier, giving us cues that the sound comes from the rear. The ILD differences from the main factors and secondary factors can create quite a significant difference from 3dB up to 20dB attenuation at higher frequencies. Binaural recordings exploit the Duplex theory to give a convincing immersive experience. To quantify the binaural quality of different binaural recording techniques, we can compare the ITD, ILD, and IPD results of binaurally recorded audio employing different methods and techniques in an experiment.

## 2.2   Interaural time differences

A Simple ITD model (Figure 2.1) for the ITD differences in two ears when a sound source is coming at an angle from the head's frontal center can be estimated with the Woodworth model. The Woodworth's ITD model involves calculating the additional travel time that a sound wavefront from a source would need to take to cover the additional distance that spans ear to ear without cutting through the head itself- meaning the sound would have to travel in a curvature around the head's arc length to reach the other ear.

For example, for a sound source that is at a 45-degree azimuth to a head's left side, we would expect that there will be an additional travel distance of $d_2 = 7\text{-}10$ cm for the sound wavefront to reach the right ear. (Figure 2.2)

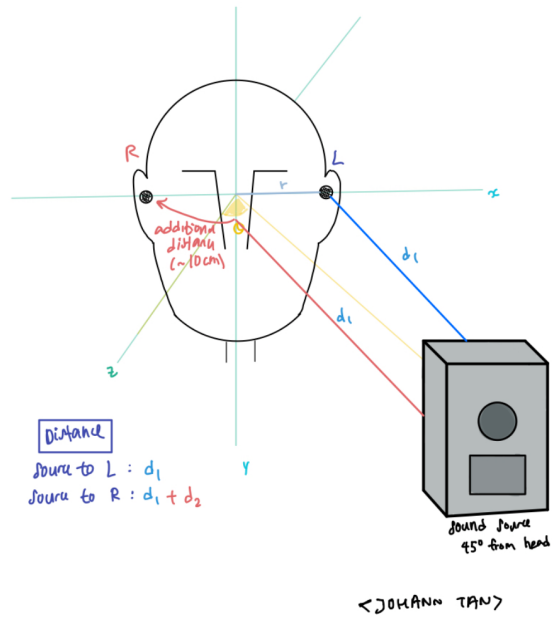

Figure 2.1: Simple ITD model front visualisation

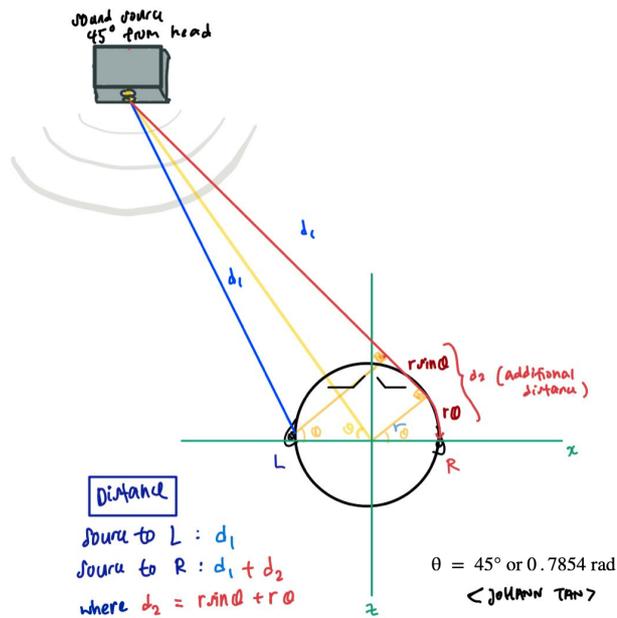

Figure 2.2: Simple ITD model top-down visualisation

From the illustration, the simple ITD model can be modeled for any individual's head radius and the azimuth angle to the sound source from the head's center. Distance to the sound source is not required and in theory, will not vastly differ for sound sources in the near field or far field from the head. (Brungart, 1998; Duda and Martens, 1998)

$$ITD_s = \frac{r(\theta + sin\theta)}{c} \tag{2.1}$$

where $r$ refers to radius in metres, $\theta$ is the angle in radians, and $c$ is the speed of sound in m/s where $c = 343m/s$ at $20°C$

However, the simple ITD model does not take into account the frequency of the sound source (unless it is a wide spectrum of frequencies). At the lower frequencies, diffraction around the head will cause a longer ITD difference. (Kuhn, 1977) The model also does not take into account slight changes in the speed of sound at different temperatures. A modified Woodworth ITD model will more accurately determine the ITD as follows;

$$ITD = \frac{ar}{331 + 0.6T}sin\theta \tag{2.2}$$

where $a = 3$ for frequencies below 500Hz and $a = 2$ for frequencies below 2000Hz.

The ITD cue for localisation is also associated with another tertiary inter-aural cue called the Inter-aural Phase Difference (IPD). (Figure 2.3) The phase difference between two ears may not be very significant, thus, it is also only useful in the lower frequency for sound localization. The out-of-phase in the wavelength of binaural content may be desired for the replication of binaural hearing.

## 2.3  Interlevel differences

It is more difficult to model the ILD exactly as it is dependent on varying angles and different frequencies. However, it has been studied experimentally that ILD at higher frequencies will be more discernible as high frequencies diffract less around the head- casting an acoustic shadow, thereby causing a significant drop in the level difference between two ears which helps us with sound localization and also contributes to the binaural effect. (Figure 2.4 & Figure 2.5)

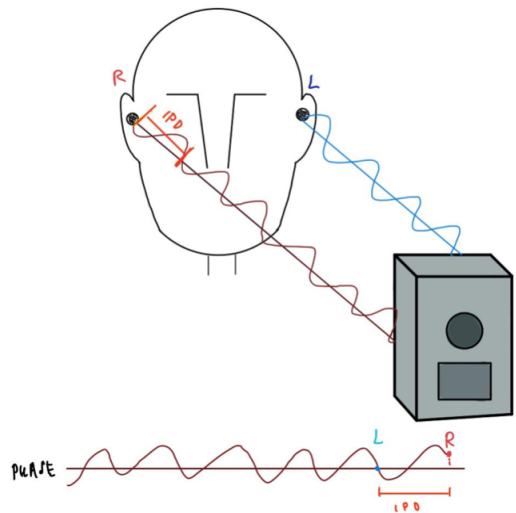

Figure 2.3: Inter-aural phase difference

## 2.4 Spectral cues

Spectral cues are monaural cues that help us with sound localization in the elevation (vertical) plane which helps us to discern sound coming from the top or the bottom. Spectral cues mainly use the shape of our pinnae which affects the sound spectrally in frequency as it enters the ears (like an EQ filter). Through the process, our brains can identify the direction and localization of sound from the spectral information captured by our ears. However, as spectral cues are not binaural cues- meaning that these spectral cues are just as effective using one ear, they are not usually considered in most binaural recording methods. Only some binaural methods that include an ear-shaped pinnae which is accurate in replicating the shape of human ears' will be able to include some monaural spectral cues. Even then, because every individual's human ears are different, there is doubt about whether the monaural spectral cues translate well to different individuals when played back with headphones.

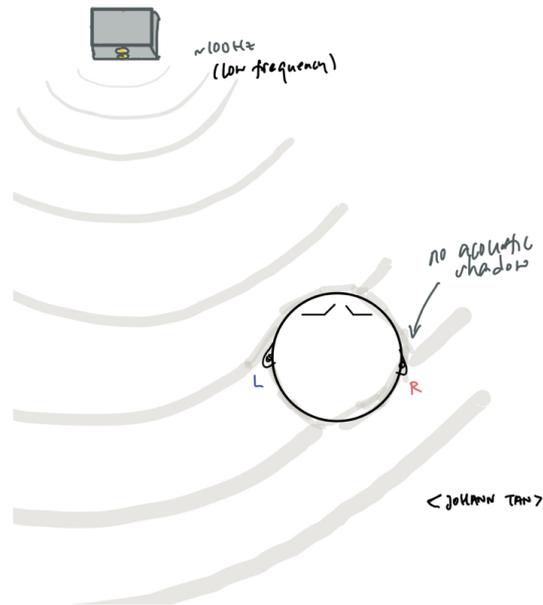

Figure 2.4: Acoustic shadow due to HRTF in low frequencies

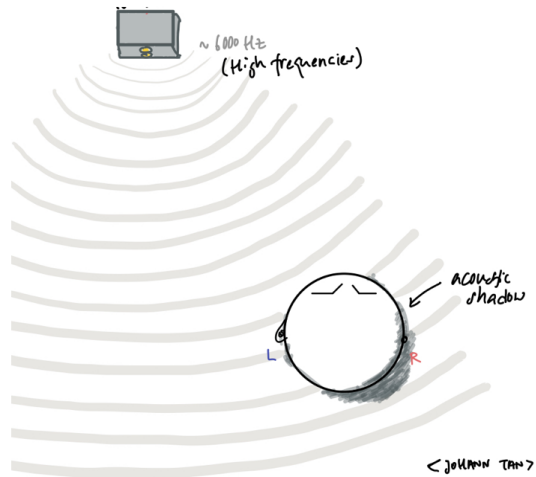

Figure 2.5: Acoustic shadow due to HRTF in high frequencies

# 3. Methodology

In this study, we compare experimental investigations between the different binaural recording methods and an actual HRTF recording in ITD and ILD with sound sources of impulse pink noise and impulse sine tones- positioned in the far-field (>3m) at the same acoustic space. They are recorded with the same omnidirectional microphones in an acoustic lab environment (but not anechoic). We will also compare the quasi-binaural ORTF technique which is considered a stereophonic recording technique as a secondary reference. The experimental testing plan schematics and laboratory environment is shown in Figure 3.1 and Figure 3.2.

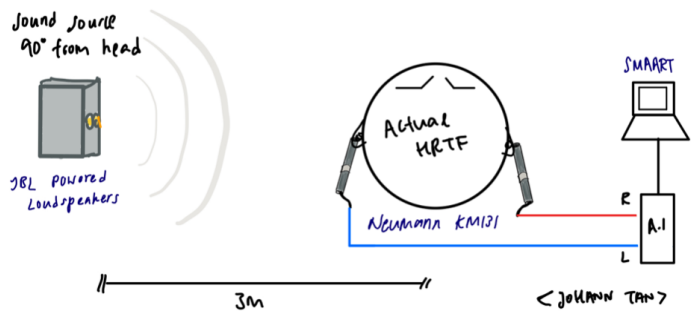

Figure 3.1: Schematics of experimental testing plan

The primary reference will be an actual HRTF by recording test signals on two sides of a subject's head with a head circumference of 56cm (r = 8.9cm). This will form a baseline for comparing binaural recording methods in their ITD and ILD results. The microphones chosen are two identical microphone models from Neumann KM 131 (Figure 3.3) as they exhibit extremely low electrical self-noise as well as little coloration and flat frequency response (Figure 3.4). The microphone also has an omnidirectional polar pattern (Figure 3.5) and is able to ignore its own presence in the recording (free-field). In other microphones, the mere presence of the microphone itself may affect the sound field it records in. Before the recording, the microphones are calibrated to the same gain level on a Scarlett USB Audio Interface. The calibration test signal (pink noise) played is generated from a Mac computer running Smaart v8 which outputs the test signal through an interface to a EON JBL Powered Loudspeaker.

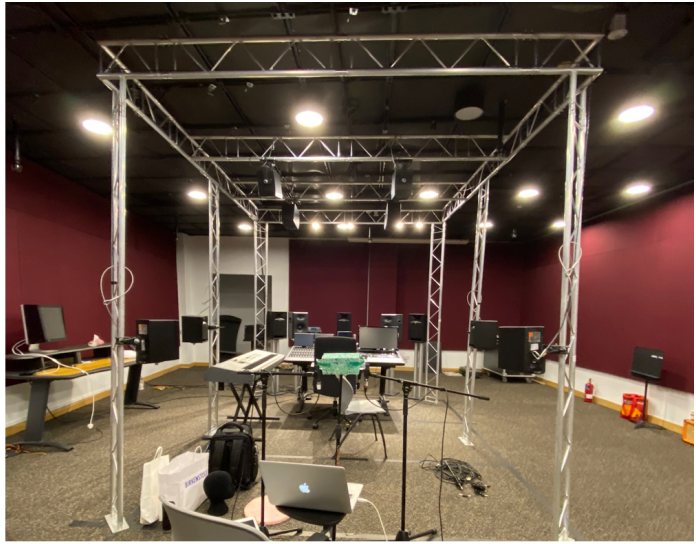

Figure 3.2: Experimental lab environment (AAS Recording lab @ YST NUS)

To calibrate the microphone, they are placed as close as possible facing each other and at the same distance to the loudspeaker while the calibration test signal is playing (Figure 3.3). The gain level and spectrum (RTA) are matched to be the same (Figure 2.1) in terms of its recorded levels and frequency response.

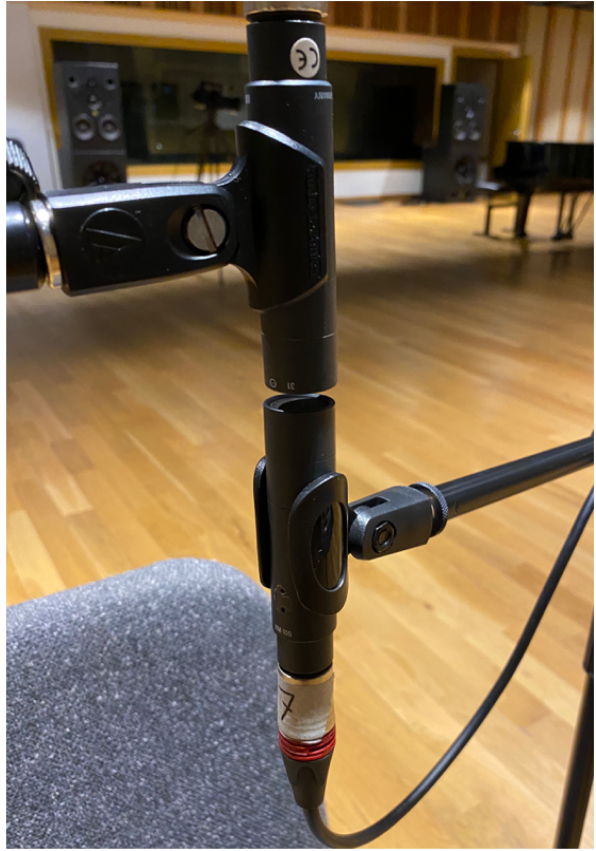

Figure 3.3: Neumann KM131 Calibration method

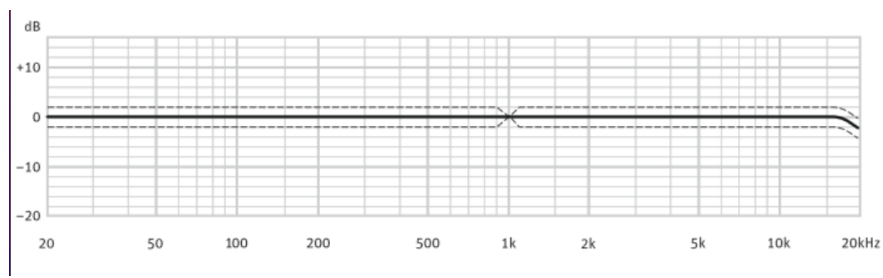

Figure 3.4: Neumann KM131 Frequency Response, Picture from Neumann

The matched RTA spectrum (Figure 3.6) does not show any deviations in the frequency response that exceeds 3dB. The next results (Figure 3.7) show the successful calibration of both receiving microphones, the transfer function's green line shows the difference between two microphones and how much phase as well as decibels they differ from one another. Since this is a calibration test, they should ideally show little to no difference in the results.

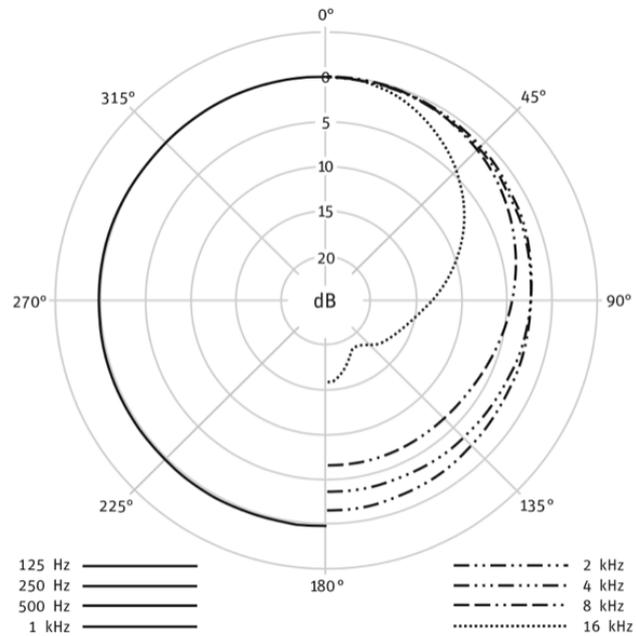

Figure 3.5:  Neumann KM131 Polar Pattern, Picture from Neumann)

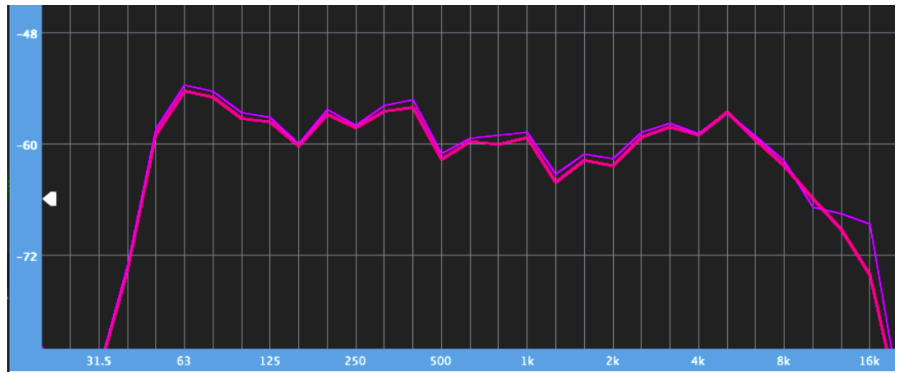

Figure 3.6:  Neumann KM131 Microphone Pair matched RTA Spectrum

The results show that the signal arrives at the same time (0 ms time difference)- which is valid, since the two microphones in calibration are occupying the same position. The two microphones show that they are generally in perfect phase with each other. The magnitude shows that the magnitude difference, dB remains at zero (0dB) until around 4kHz, where the magnitude difference starts to vary. However, it is within -3dB range tolerances between the two calibrated microphones until above 16kHz, where the magnitudes start to differ more greatly. Magnitude differences in the higher limit of frequencies become harder to control, which may be due to room acoustics or the physical microphones themselves interfering with each other due to being placed in close proximity to each other.

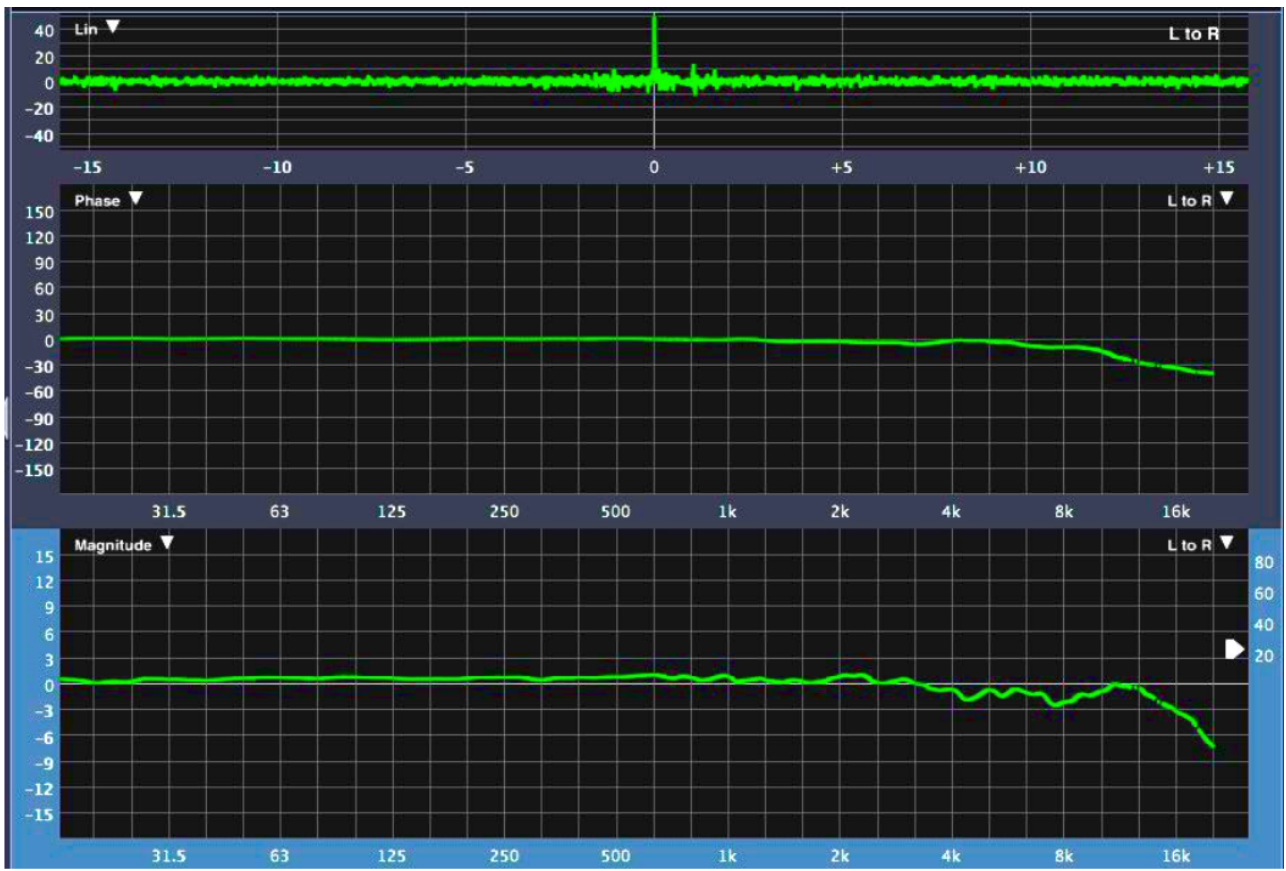

Figure 3.7: Neumann KM131 Microphone Pair Transfer functions

At 8kHz, the narrow band difference of <3dB might be due to a minute difference in the manufacturing quality of the microphone even though they are identical models. This deviation at 8kHz is adjusted and corrected for in Smaart. The Smaart v8 software is a real-time fast Fourier transfer (FFT) analysis tool that will be used to capture the experimental results of the levels, time delay as well as phase information between the two microphones using dual-FFT audio signal comparison (transfer function) while comparing between different binaural recording methods. The experiment test signal (pink noise as well as sine tones) will be generated from the Smaart software. The reference channel will be the left Neumann KM131 microphone while the measurement channel will be the right Neumann KM131 microphone. The results will show the magnitude difference (ILD) across the audible frequency spectrum at the bottom, the phase difference at the middle (IPD), and the time difference at the top (ITD).

# 4. Results & Discussions

## 4.1 Individual's HRTF

A HRTF baseline is captured with the pair of microphones placed on the sides of an individual's head (the author) as close as possible to the ears with the experiment test signal (pink noise) impulse played from a loudspeaker at 90° to the left of the individual at 3m away. The measured ITD was found to be 0.69ms, which is close to the theoretically calculated ITD from the Simple ITD model, given the known radius, $r = 0.89m$, $\theta = 90°$ and $T = 18°C$.

$$ITD_s = \frac{0.89(1.57 + sin(1.57))}{331 + 0.6(18)} = 0.669ms \tag{4.1}$$

The results from Figure 4.1 also show that ILD difference in magnitude starts to occur around 250Hz at JND decimal differences <3dB but starts to pick up after 1kHz, reaching -6dB at 2kHz in level differences between the left and right ear microphones. After 4kHz, the roll-off becomes significantly greater than -10dB and continues to roll off to -15dB above 8kHz similar to that of a low pass filter. (see figure 28) The results are similar to the experimental findings of ILD across different frequencies noted by other studies (Figure 4.2) and thus form a good baseline to follow in comparison with the different binaural recording methods in the replication of human hearing.

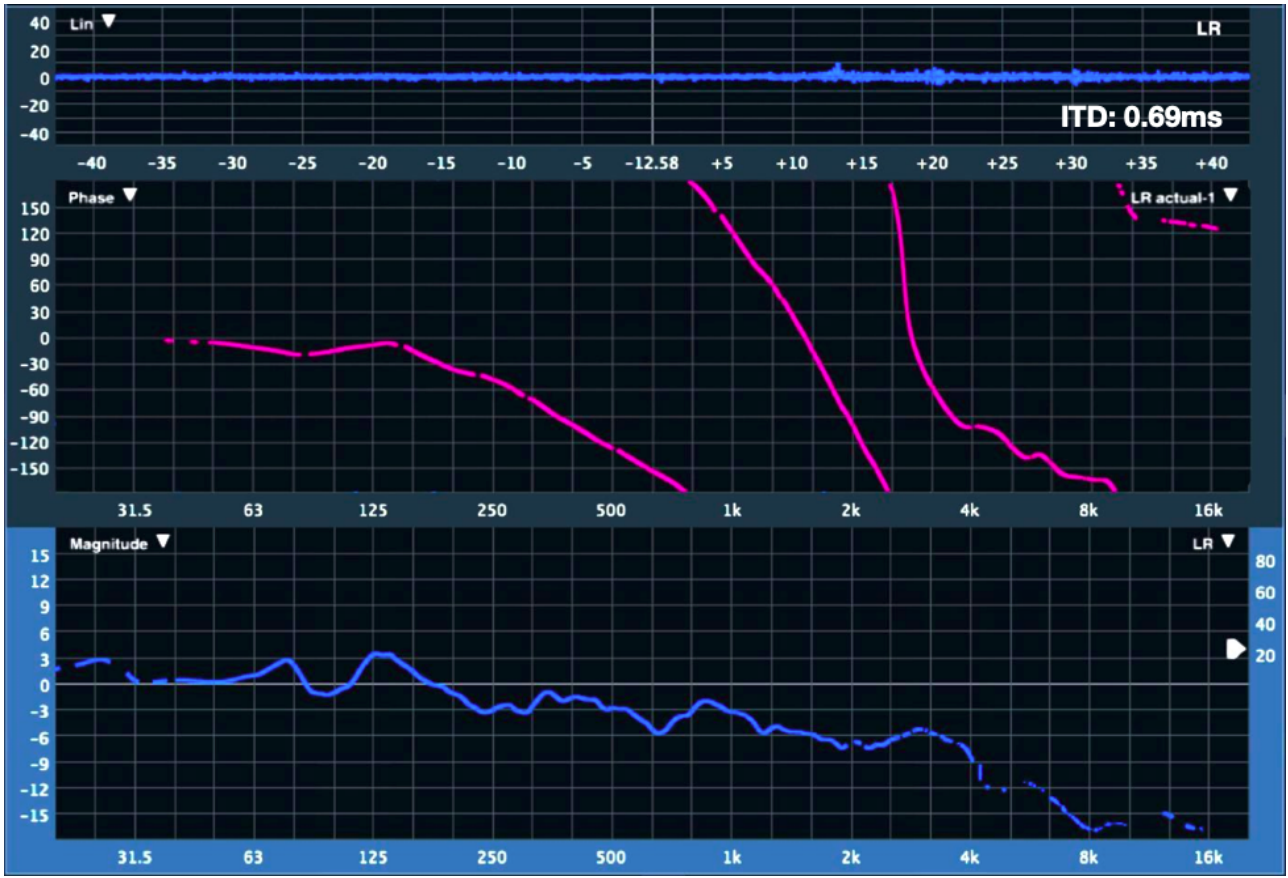

Figure 4.1: An individual's head-related transfer function - Time, phase, and magnitude measurement. ITD = 0.69 ms

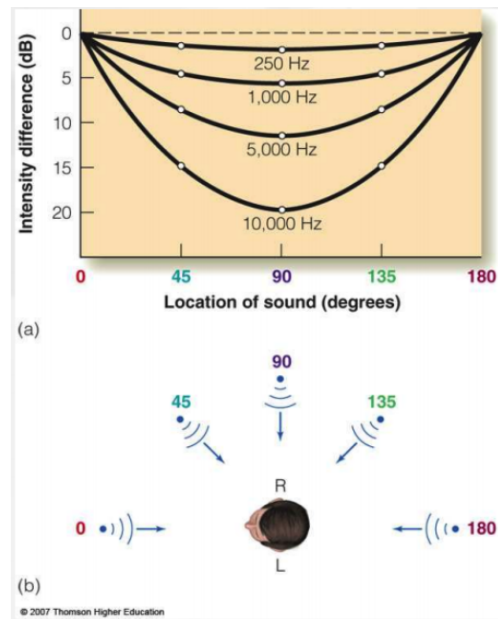

Figure 4.2: Interlevel differences for different frequencies and angles, Picture from Thomson Higher Education

## 4.2 Full dummy head

The measurement for the full dummy head binaural method was captured by placing the same microphone on the sides of a dummy head made with PVC material. The choice of PVC material for the dummy head was to achieve a close absorption coefficient to that of the human head. The ITD was measured to be 0.67ms which is 0.03ms faster than the HRTF ITD. The magnitude difference shows an attenuation of 3 to 6dB from 500Hz to 8kHz between the left and the right microphone for the experiment test signal which is directed from the same position as the previous test. This drop remains consistent and then decreases steeply from 11kHz onwards. (Figure 4.3)

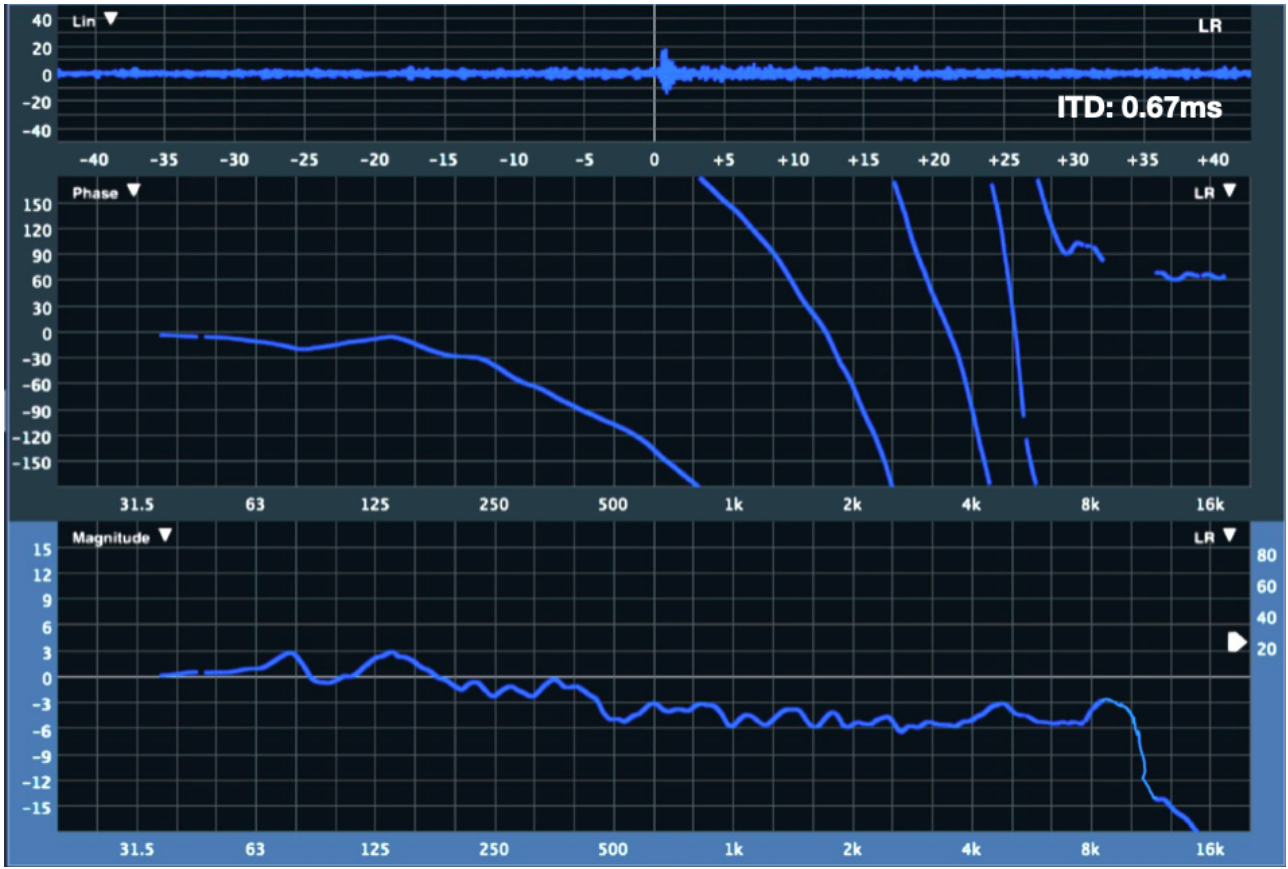

Figure 4.3: Full dummy head's - Time, phase, and magnitude measurement. ITD = 0.69 ms

### 4.3 Semi dummy head

The measurement for the semi-dummy head binaural recording method was captured by placing the same microphone on the sides of a pair of silicon ear models spaced 19cm apart with a melamine foam-like material acting as an obstruction between the ear models. The ITD was measured to be 0.83ms. The magnitude results show minimal ILD <3dB in the low to high mid frequencies and only start to show noticeable ILD from 2kHz onwards which then drops significantly and promptly from 4kHz. (Figure 4.4)

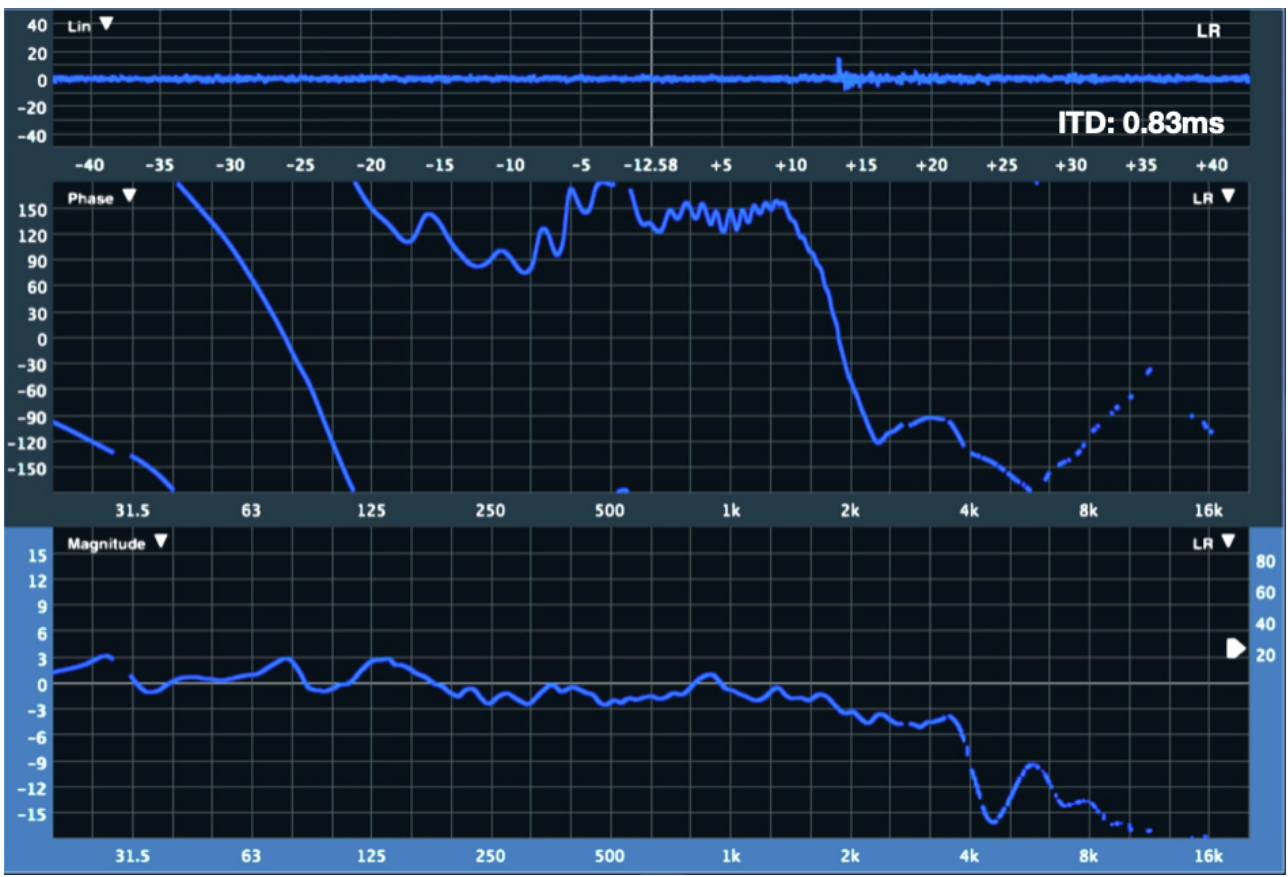

Figure 4.4: Semi dummy head's - Time, phase, and magnitude measurement. ITD = 0.83 ms

### 4.4   Jecklin Disc

The measurement for the Jecklin disc acoustic baffle method was captured with the same microphone on the two sides of the Jecklin disc placed 17.5cm apart. The Jecklin disc acts as an obstruction between the two microphones and is made from a foam material that is cut to 33cm in diameter. The ITD measured was 0.58ms. The magnitude results show minimal ILD <3dB across the entire spectrum except from 6kHz onwards. However, the ILD is not markedly pronounced as compared to previous measurements with 8dB difference across the high-frequency range. Interestingly, this result is the only one that doesn't have a large and steep drop in the very high frequencies >10kHz but this is not necessarily favorable. (Figure 4.5)

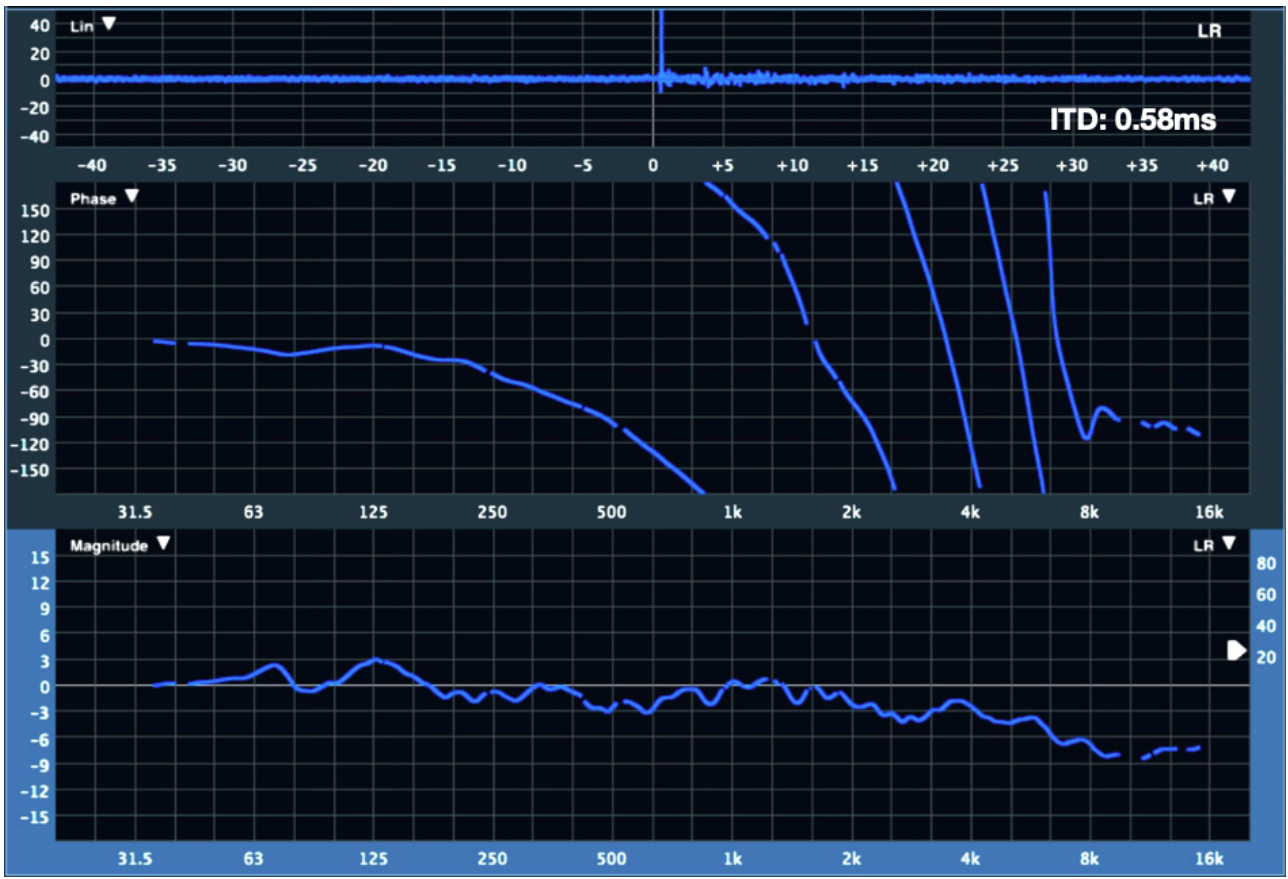

Figure 4.5: Jecklin disc's - Time, phase, and magnitude measurement. ITD = 0.58 ms

## 4.5 ORTF

The measurement for ORTF technique follows the stereophonic recording technique that can produce a quasi-binaural image. The pair of microphones is placed with the capsule spaced 17cm apart and angled at 110°. The ITD measured was 0.50ms. The magnitude shows minimal ILD <3dB in the low and mid-high frequencies. From 4kHz onwards, the results show a smooth downward slope akin to a low pass filter. (Figure 4.6)

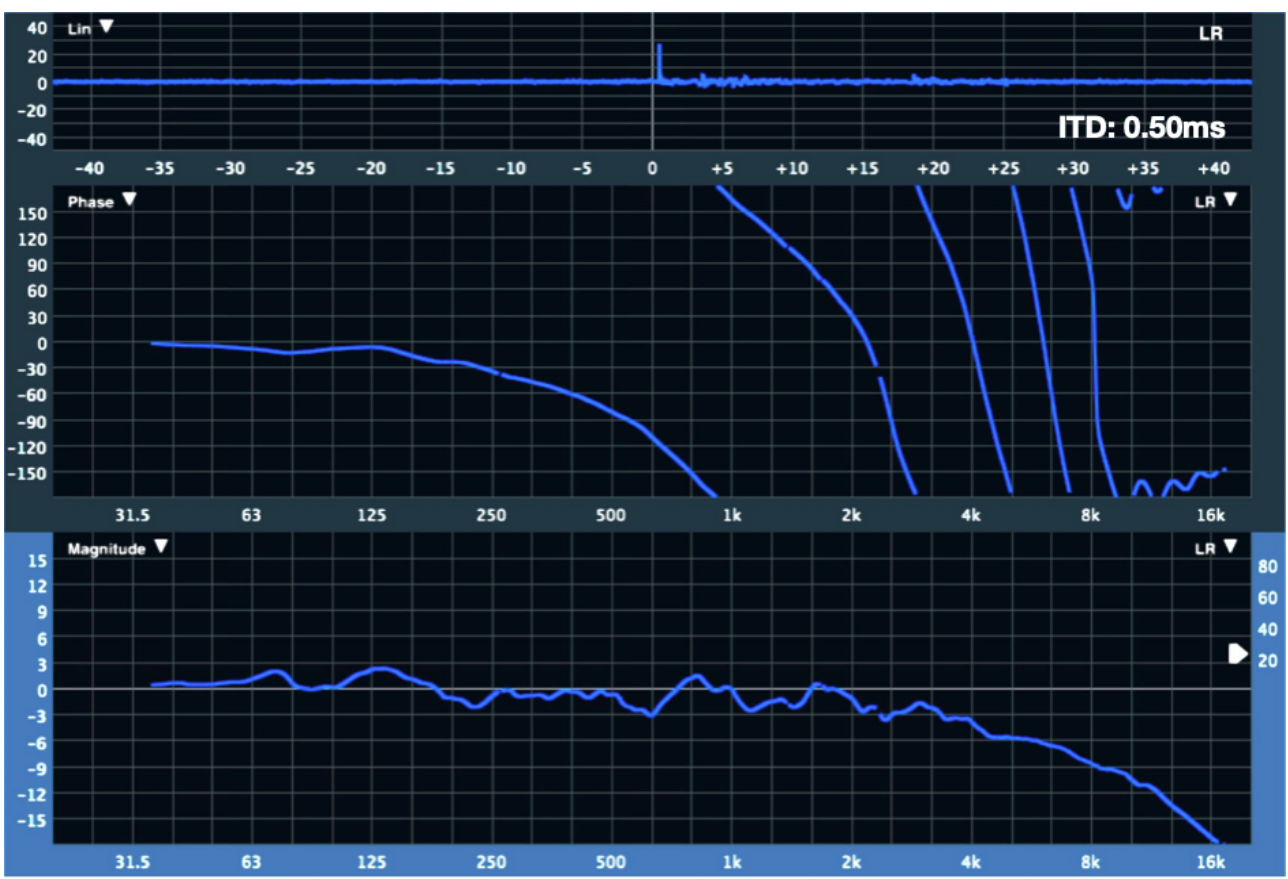

Figure 4.6: ORTF Technique - Time, phase, and magnitude measurement. ITD = 0.50 ms

### 4.6   Interlevel differences comparisons to baseline

Based on the observations and the comparison chart from Figure 4.7 and Figure 4.8, it can be shown that the result of the HRTF Baseline is closest to the result of the Full Dummy Head which would be within our expectations.  However, from 4kHz onwards, the Full Dummy Head as well as all the other binaural recording methods do not follow the same downward slope as the HRTF Baseline except for the semi-dummy head.  (Figure 4.7) A high cut equalization might help the Full Dummy Head achieve a closer binaural image to that of the HRTF Baseline in the 4kHz frequency range onwards.

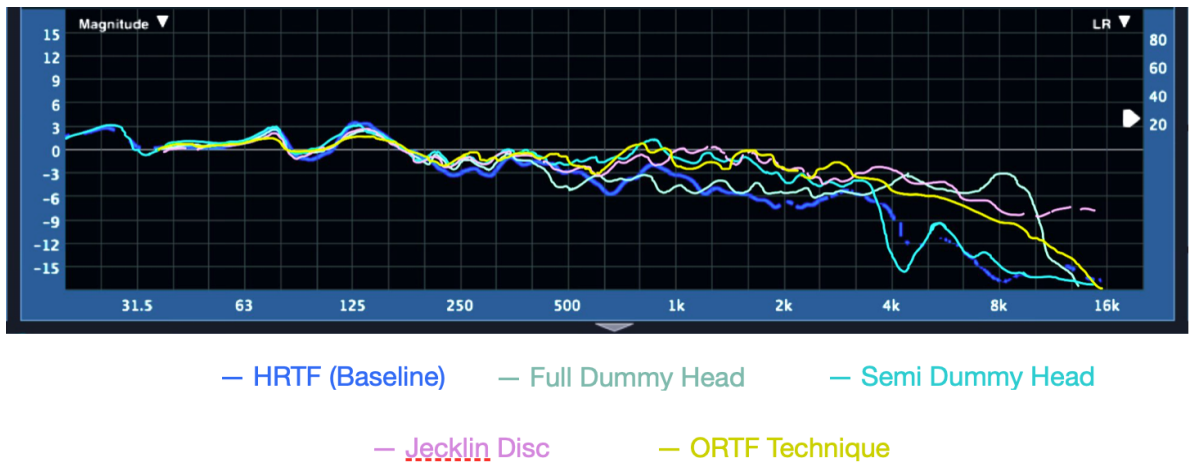

Figure 4.7: Interlevel differences comparisons for all recording methods

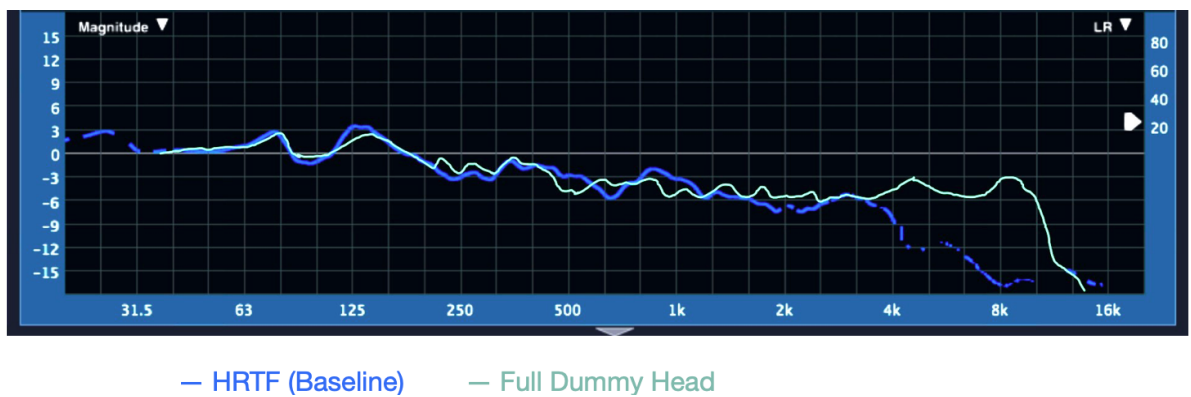

Figure 4.8: Interlevel differences comparisons between HRTF baseline and full dummy head

The HRTF Baseline and Semi dummy head results show a more aligned high-frequency response, however, it has a null at about 4.2kHz. (Figure 4.9) This might be due to an artifact from the foam material used for the semi dummy head binaural recording method.

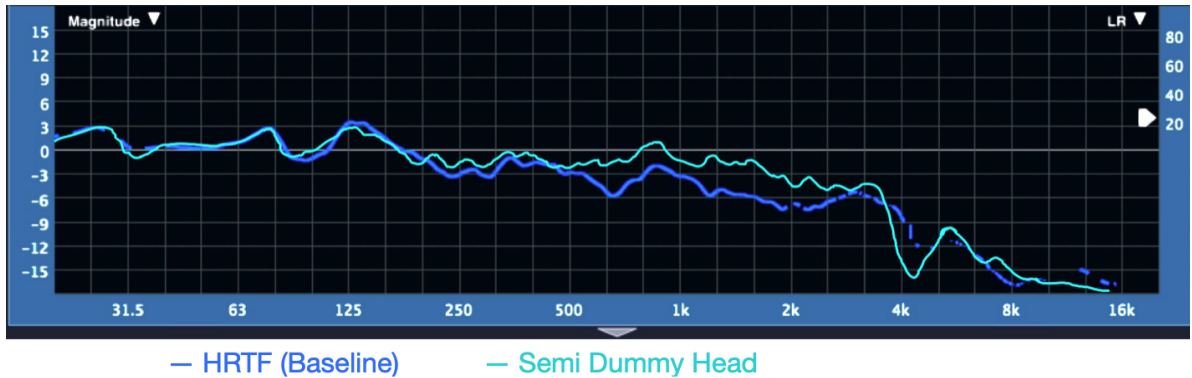

Figure 4.9: Interlevel differences comparisons between HRTF baseline and semi dummy head

The Jecklin disc and ORTF technique results show the least similarity to the binaural HRTF Baseline frequency response. (Figure 4.10 & Figure 4.11) The Jecklin disc response is more similar to the QRTF technique as a quasi-binaural approach.

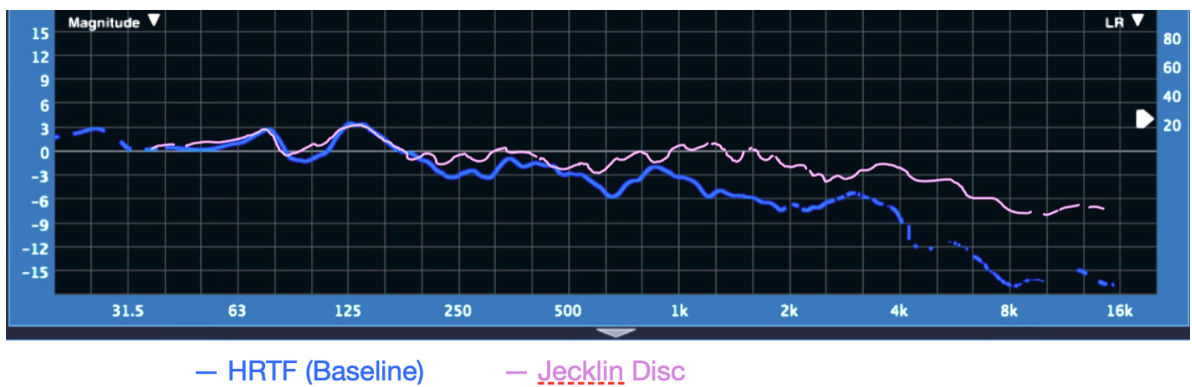

Figure 4.10: Interlevel differences comparisons between HRTF baseline and Jecklin disc

Based on the ILD comparisons, the full dummy head and semi-dummy head approach appears to be more suitable as a binaural recording method as opposed to the Jecklin and ORTF techniques. The Full dummy head would be a perfect choice if the frequency response above 4kHz matches that of the HRTF Baseline. A better result that extends to the higher frequencies may be achievable with a more absorptive material like that of the melamine foam material used for the Semi-Dummy head instead of PVC. However, since such a foam dummy head is not readily available- other alternatives may work. Another approach would be EQ adjustments post-recording to further attenuate the higher frequencies to achieve a binaural frequency response that more closely resembles the baseline HRTF results.

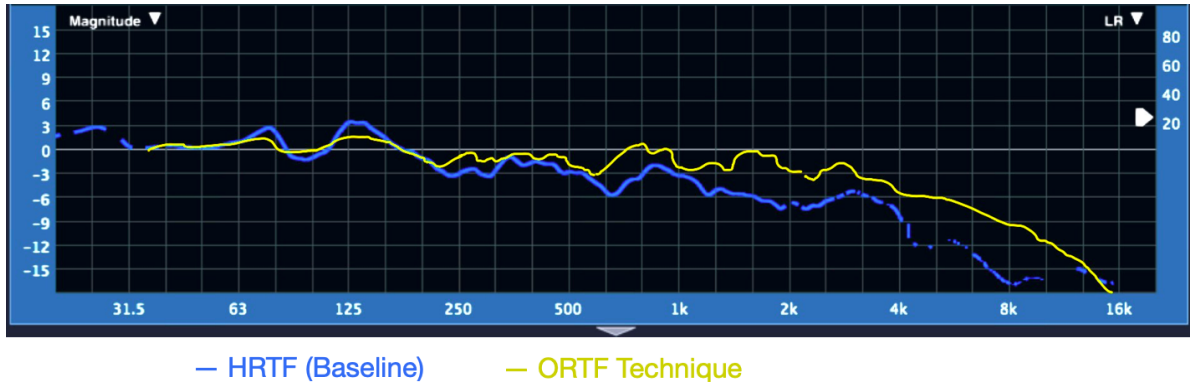

Figure 4.11: Interlevel differences comparisons between HRTF baseline and ORTF technique

## 4.7   Intertime difference comparisons to baseline

The ITD measured for the HRTF Baseline was 0.69ms, the only ITD results that matched the HRTF Baseline was the Full dummy head. For the Semi-dummy head, the microphones were spaced much further apart than the typical distance that separates the human ears. A larger difference in ITD is to be expected (0.83ms). This means that the Semi-dummy head will be less effective in the localization of lower-frequency sounds such as the low rumble of engines or approaching cars. Reducing the distance between the semi-dummy heads would decrease the ITD to more closely match the HRTF Baseline's ITD. Future consideration may be to explore a shorter Semi-dummy head contraption and determine if this would compromise the binaural effect in other aspects. (Such as the ILD and IPD) Figure 4.12 shows the results and comparisons of all the ITD measurements.

During the ITD experimentation, it was also observed that the ITD between the left and right microphones was different depending on the frequencies used. (Figure 4.13) In the

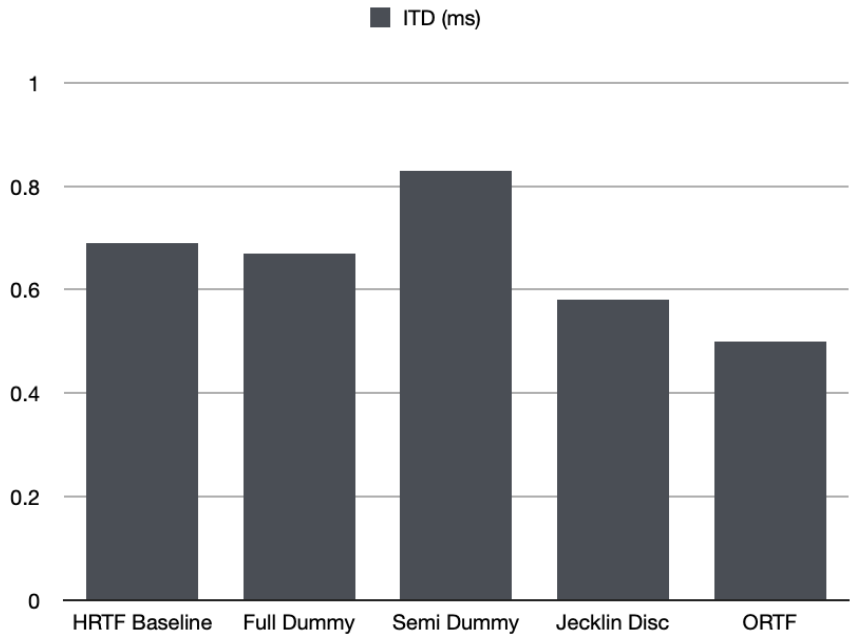

Figure 4.12: Interaural time difference comparisons (Pink noise)

pink noise test signal experiment, it was observed that the ITD would show a time value from the onset of the pink noise received from the microphones as the sound travels from the left to the right microphone, the slight delay between the left and right microphone gives the ITD value. However, since the pink noise is a full spectral of frequencies, the higher frequencies likely influence the ITD value more than the lower frequencies. Since the lower frequencies will take a slightly longer travel time. (Kuhn, 1977) Figure 4.13 shows the ITD values based on the sine frequencies that were used (Low : 220Hz and High : 6kHz)

It can be observed that the HRTF Baseline has about $42\mu s$ (0.042ms) ITD difference when compared to low and high frequencies ITD values. The same ITD difference was observed in the Full Dummy Head. However, the Semi Dummy Head and ORTF- both had greater ITD differences between the low and high frequencies time delay. (Semi Dummy: $80\mu s$, ORTF: $83\mu s$) This difference in the time delay between the low and high frequency is due in part to the distance between the left and right microphone as well as the minute difference in diffraction pattern as the low-frequency content is transmitted over and around the head in different ways between the various binaural recording setups. In considering the pink noise ITD and sine tones tests, the Full dummy head binaural method would be the most optimal for recording binaural to match that of the Human HRTF (Individualised).

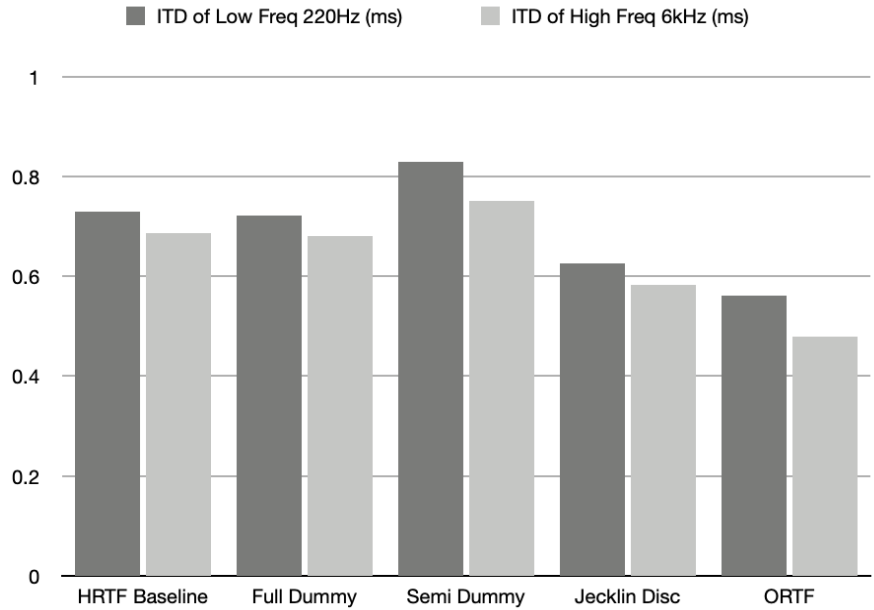

Figure 4.13: Interaural time difference comparisons (Sine signal)

Taking into account both the experimental ILD and ITD measurements of the various popular methods utilized for binaural recordings. The most accurate representation of the human HRTF (Individualised to the subject's HRTF) is the Full Dummy Head method and the Semi Dummy Head method. While the rest of the binaural recording methods significantly differed from the Individualised HRTF results.

# 5. Limitations

## 5.1 Cone of confusion from static head

An important aspect of sound localization is the movement of our heads to detect where ambiguous auditory sources are coming from- these ambiguous auditory sources are usually from a region of space around a person's head that is called the 'cone of confusion' (Figure 5.1) where our ears are unable to determine correctly the localisation of the auditory source. This is because auditory sources from any points along this cone of confusion have the same apparent ITD and ILD even though they are located at entirely different positions. In reality, the cone of confusion can be easily resolved with head movements such as facing to the left or right to determine the actual position of auditory sources. However, all the methods of binaural recording involve a static head that cannot be moved or turned. The localization of sound and the binaural effect are limited by the cone of confusion.

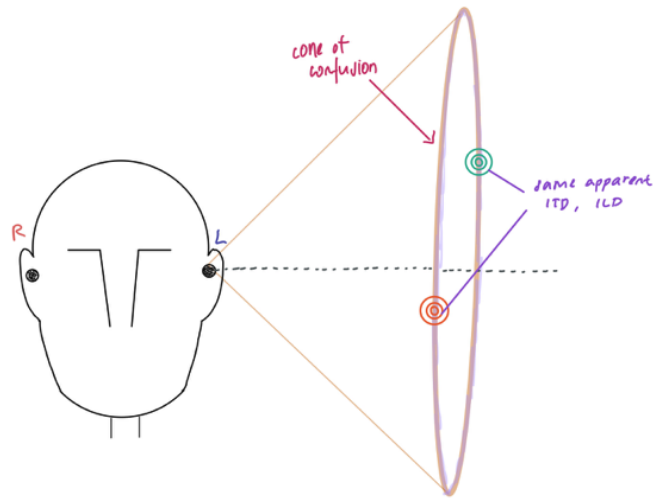

Figure 5.1: Cone of confusion

## 5.2   Individualised HRTF and Spectral cues due to the ear's pinnae

As previously discussed, because of the individual size of each individual's HRTF as well as pinnae shape. The difference in HRTF may not translate with 100% authenticity to different individuals listening to the playback of the binaural audio, which depends on the similarity in HRTF of the individual's head to the binaurally recorded HRTF. While this limitation exists, it will not affect the binaural effect to the extent that it cannot be used for its immersion. Monaural spectral cues that involve the ear's pinnae were also not represented well due to the difference in the shape of the pinnae. Frequencies around 2kHz - 3kHz would be most affected by the exclusion of the pinnae in the binaural recordings. (Marko Tapio Hiipakka, 2008) It should also be noted that the individual's listening to binaural content is also dependent on the individual's hearing health and preferences.

## 5.3   Use of binaural content in virtual reality applications

For some time, binaural audio content has been useful in VR applications, however with the introduction of VR headset and 360° videos, binaural audio presents a problem due to its static nature that does not account for head movement. In VR, the viewer's head's movement is tracked and the video moves together with the motion of the head. However, current binaural audio technology lacks the capability to track the head and follow the perspective of the viewer's ears. Which creates a disconnect in the video and audio as the viewer turns his head. As the audio can only be pre-mixed to follow a single direction/perspective in which most cases would be the forward perspective in the aural mix. This limitation can be resolved with a more advanced recording method- such as the complex Higher Order Ambisonics (HOA) technique which is of at least 2nd order ambisonics used for recording 360° spatial audio that can later be tracked to the head movement of the user. This technology is still in its infancy stage but shows great potential in VR applications.

## 5.4   Microphone's frequency response

The microphone that was used had a frequency response that was entirely clean and flat. (Figure 5.2) Thus, there was no coloration of sound spectrum across the frequencies which is not representative of how our ears hear in regards to spectral information, according to a study conducted by Marko Hiipakka, 2008. (Figure 5.3) A flat frequency response is not the best representation of natural human hearing. The way that our ear canal shapes the spectral structure of sounds which alters auditory information has not been fully studied and is not completely known. Some commercially available binaural recording microphones may have the microphone's frequency response tweaked to create more naturally sounding recordings that follow how our ears hear frequency spectral information.

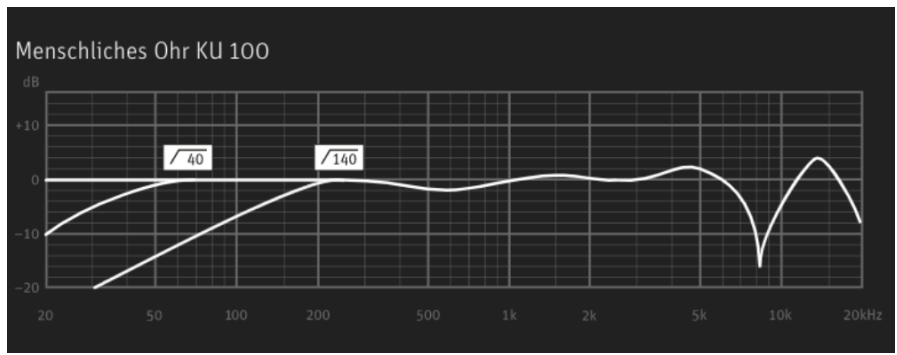

Figure 5.2: Frequency response of the Neumann KU 100 microphone with a dip around 8kHz (Picture from Neumann)

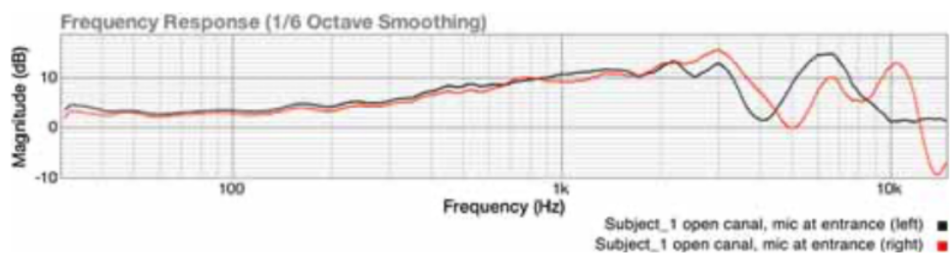

Figure 5.3: Frequency response of actual human ears (Picture from Marko Hiipakka)

## 5.5    Material's absorption coefficients

The human head's sound actual absorption coefficient, $\alpha$, would affect the absorption of sound differently when compared to other materials. The difference between the materials used for each binaural recording method would slightly affect the results. However the exact difference is not known. The human body is around $\alpha = 0.2 - 0.5$ whereas the PVC material used for the dummy head has a typical absorption value of $\alpha = 0.4 - 0.9$. The semi-sized dummy head as well as the Jecklin disc made from foam material also has different sound absorption coefficient values in the very high frequency ranges.

# 6. Other Considerations

## 6.1   Binaural re-recordings

A similar technique to Worldizing, which is a method where a sound effect can be re-recorded and reproduced in a different environmental setting, and microphones can also be applied to non-binaural audio and create binaural content from them. This re-binauralisation method can be used to record multi-channel speaker setups such as a surround system and convert a surround playback into a binaural playback. However, the technique for re-binauralisation is still not yet fully established.

## 6.2   Binaural Virtual Studio Technology

Binaural spatialization virtual studio engine (VST) like the Anaglpyh toolbox- developed by David Poirier-Quinot opens up the possibility for sound engineers to synthesize binaural effects from mono or stereo audio recordings. The engine uses the results of many years of spatial hearing research involving binaural hearing. The engine is in its beta release with ongoing binaural research and is only currently compatible with a few audio workstations like Ableton and Reaper on Macintosh computers. The Anaglpyh engine also involves the ILD and ITD that we have discussed previously in its binaural rendering as well as HRTF parameters. It also lets the user customize the individual's HRTF into the binaural rendering. All of this contributes to the binaural rendering of sound to make binaural convincing even though the original recording may be from a monaural source. Another binaural spatialisation VST is the Ambeo Orbit by Sennheiser. Although it is more compatibility with different DAWs, it is not as comprehensive in its binauralisation effect as the Anaglpyh VST.

## 6.3   Binaural Mixing

When it comes to mixing for binaural audio that involves binauralisation techniques, it is best to avoid mixing in loudspeaker playback as the binaural effect will be lost from multiple reflections in an acoustical room. Even with transauralisation configuration, the use of headphones for mixing binaural content presents a lot more advantages than loudspeakers, after all the end user is expected to listen to the binaural content with headphones. A short verification check on the playback of binaural audio on loudspeakers may be the only reason for mixing via loudspeakers. Regarding the use of headphones, open-backed headphones is more suitable than closed-back headphones for binaural mixing as closed-back headphones tend to emphasize the lower frequency and create internal resonance during playback. An unwanted effect of listening in closed-back headphones is that it would represent sound sources closer than they actually are due to the emphasis on lower frequencies.

## 6.4   Binaural Headset or In-Ear Microphone Recording Method

The experimental measurement of the ITD and ILD results of the Binaural headsets/in-ear microphone was not done since the experimental microphone used (Neumann KM131) could not fit as an in-ear microphone. Hence, this binaural recording method was not included in the experimentation. Although the HRTF method would give a good estimation of the performance of the in-ear microphone recording method. The microphones used for such devices tend to be small and built into the headset or in-ear itself. Thus, the frequency response would vary, although it is expected that the ITD will not be different. Since the microphone used is not the same, the results shown below (Figure 6.1) are the ILD difference between the left and right HRTF results when recorded with a binaural headset, which is not equivalent in terms of quality as compared to the studio quality condenser microphones and hence was not included in the experimental discussion.

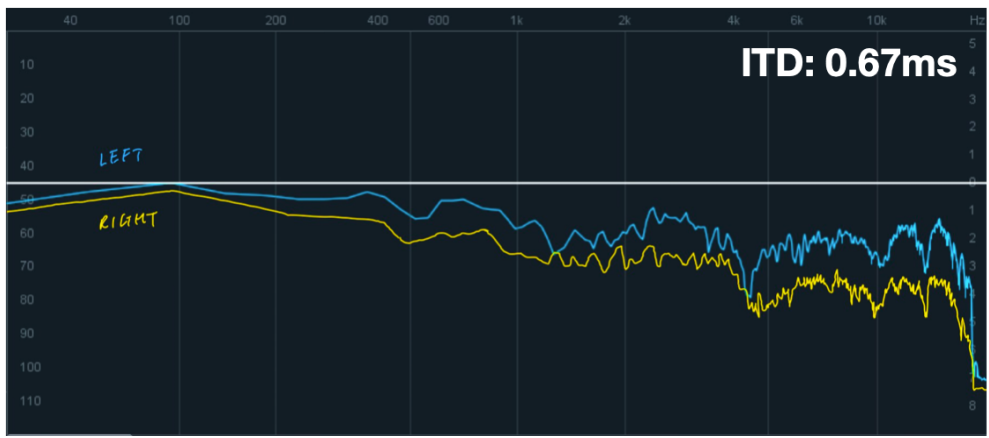

Figure 6.1: Individualised HRTF, recorded with binaural headset microphone (Sennheiser Ambeo Headset)

# 7. Appendix (Documented photographs)

HRTF Baseline Photo Documentation

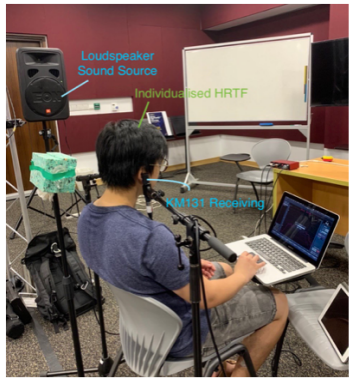

*Figure A1. Individualised HRTF Measurement*

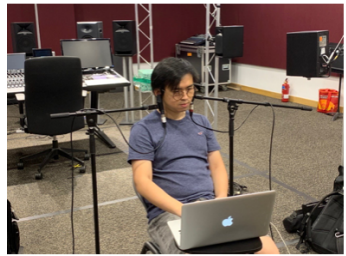

*Figure A2. Individualised HRTF Measurement*

Artificial Full Dummy Head

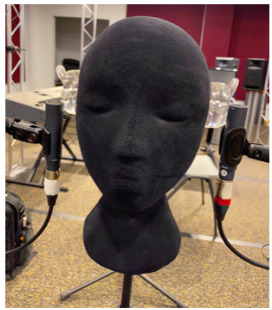

*Figure A3. Full Dummy Head Measurement*

Artificial Semi Dummy Head

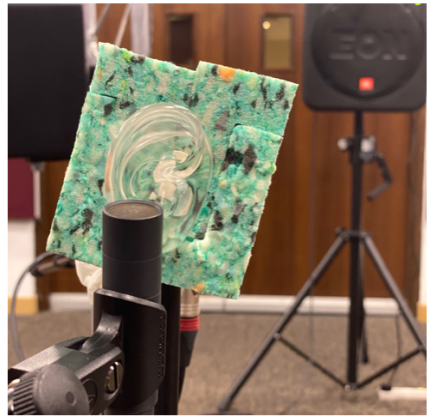

*Figure A4. Semi Dummy Head Measurement*

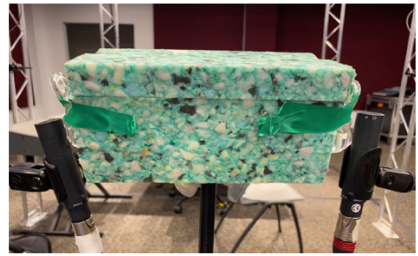

*Figure A5. Semi Dummy Head Measurement*

<u>Jecklin Disc (Baffled Stereo)</u>

ORTF Stereophonic Technique

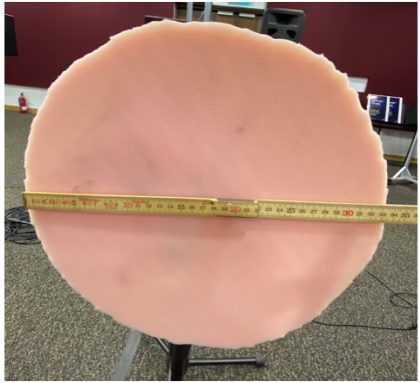

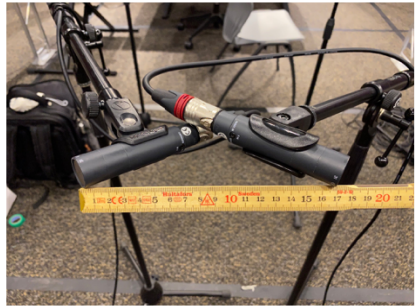

*Figure A6. Jecklin Disc Measurement*

*Figure A8. ORTF Measurement*

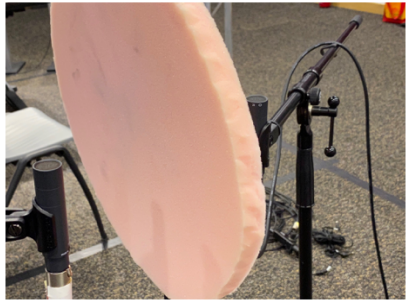

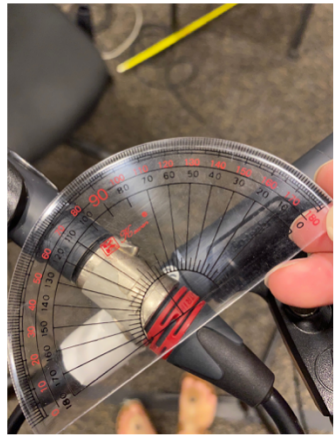

*Figure A7. Jecklin Disc Measurement*

*Figure A9. ORTF Measurement*